\documentclass[nofootinbib,aps,twocolumn,letterpaper,superscriptaddress,showpacs,prd]{revtex4}
\usepackage{mathrsfs,amsmath,amsthm,latexsym,amssymb,amsfonts,epsfig,txfonts}
\newcommand{\makeSymbol}[1]{\mathord{\vcenter{\hbox{#1}}}}
\usepackage{hyperref}
\hypersetup{bookmarks=true, bookmarksnumbered=true,
bookmarksopen=true, colorlinks, linkcolor=red, citecolor=magenta,
plainpages=false, pdfstartview=FitH}
\usepackage{booktabs}
\usepackage{verbatim}
\allowdisplaybreaks
\begin{document}

\title{{\sf New Hamiltonian constraint operator for loop quantum gravity}}
\author{Jinsong Yang}\thanks{yangksong@gmail.com}
\affiliation{Department of Physics, Guizhou university, Guizhou 550025, China}
\affiliation{Institute of Physics, Academia Sinica, Taiwan}
\author{Yongge Ma}\thanks{Corresponding author: mayg@bnu.edu.cn}
\affiliation{Department of Physics, Beijing Normal University, Beijing 100875, China}

\begin{abstract}
A new symmetric Hamiltonian constraint operator is proposed for loop quantum gravity, which is well defined in the Hilbert space of diffeomorphism invariant states up to non-planar vertices with valence higher than three.  It inherits the advantage of the original regularization method, so that its regulated version in the kinematical Hilbert space is diffeomorphism covariant and creates new vertices to the spin networks. The quantum algebra of this Hamiltonian is anomaly-free on shell, and there is less ambiguity in its construction in comparison with the original method. The regularization procedure for this Hamiltonian constraint operator can also be applied to the symmetric model of loop quantum cosmology, which leads to a new quantum dynamics of the cosmological model.
\end{abstract}

\maketitle

The singularity theorem of general relativity (GR) is a strong signal that the classical Einstein's equations can not be trusted when the spacetime curvature grows unboundedly. It is widely expected that a quantum theory of gravity would overcome the singularity problem of classical GR. A very lesson that one can learn from GR is that the spacetime geometry itself becomes dynamical. To carry out this crucial idea raised by Einstein 100 years ago, loop quantum gravity (LQG) is notable for its nonpertuibative and background-independent construction \cite{Thiemann:2007bk,Rovelli:2004tv,Ashtekar:2004eh,Han:2005km}. The kinematical Hilbert space of LQG consists of cylindrical functions over finite graphs embedded in the spatial manifold. The quantum geometric operators corresponding to  area \cite{Rovelli:1994ge,Ashtekar:1996eg}, volume \cite{Rovelli:1994ge,Ashtekar:1997fb,Thiemann:1996au}, length\cite{Thiemann:1996at,Bianchi:2008es,Ma:2010fy}, ADM energy \cite{Thiemann:1997rs} and quasi-local energy \cite{Yang:2008th}, etc. have discrete spectrums. The LQG quantization framework can also be generalized to high-dimensional GR \cite{Bodendorfer:2011nx} and scalar-tensor theories of gravity \cite{Zhang:2011vi,Zhang:2011vg}. A crucial topic now in LQG is its quantum dynamics, which is being attacked from both the canonical LQG and the path integral approach of spin foam models. In the canonical approach a suitable regularization procedure was first proposed by Thiemann to obtain well-defined Hamiltonian constraint operators \cite{Thiemann:1996aw}. The Hamiltonian constraint operators obtained in this way will attach new arcs (edges) and hence create new trivalent co-planar vertices to the graph of the cylindrical function which it acts \cite{Thiemann:1996aw,Gaul:2000ba}. The quantum dynamics determinatted by the Hamiltonian constraint operator is well tested in the symmetric models of loop quantum cosmology (LQC) \cite{Ashtekar:2003hd}. The classical big bang singularities are resolved by quantum bounces in the models \cite{Ashtekar:2006wn,Ding:2008tq,Yang:2009fp}. However, there are ambiguities in the graph-dependent triangulation construction of this operator. In order to obtain the on shell anomaly-free quantum algebra of the Hamiltonian constraint operator  \cite{Gambini:1997bc}, one has to employ degenerate triangulation at the co-planar vertices of spin networks in the regularization procedure of the Hamiltonian.\footnote{Thanks to the remark from Thomas Thiemann.} This treatment implies that the regularization procedure has essentially neglected the Hamiltonian at the co-planar vertices before acting the regulated operator on them. Otherwise, this kind of Hamiltonian constraint operator would generate an anomalous algebra in the full theory, unless one inputs certain unnatural requirement to the interaction manner of the edges of the graph and the arcs added by the Hamiltonian operator \cite{Thiemann:1996av}. The Hamiltonian constraint operators proposed recently in \cite{Lewandowski:2014hza,Alesci:2015wla,Assanioussi:2015gka} do not generate new vertices on the graph of the cylindrical function and hence are anomaly-free on shell. However this kind of action can not match the quantum dynamics of spin foam models where new vertices are unavoidable in their construction \cite{Perez:2012wv}.  A regularization of the Hamiltonian constraint compatible with the spinfoam dynamics was considered in \cite{Alesci:2010gb}. However, the resulted Hamiltonian operator acts non-trivially on the vertices that it created and thus has still an anomalous quantum algebra. It is therefore natural to ask the question whether one can construct some Hamiltonian constraint operator with the following properties: (i) it is diffeomorphism covariant, symmetric and anomaly-free; (ii) it generates new vertices; (iii) its action on co-planar vertices is not neglected by some special regularization procedure, and there is no special restriction on the interaction manner of the edges of the graph and the arcs added by its action. We will show that the answer is affirmative. An alternative quantization of the Hamiltonian constraint in LQG possessing the above three properties will be proposed. The regularization procedure of the Hamiltonian operator can also be applied to LQC models.

The Hamiltonian formalism of GR is formulated on a 4-dimensional manifold $M=\mathbb{R}\times \Sigma$, with $\Sigma$ being a 3-dimensional spatial manifold. In connection dynamics, the canonical variables on $\Sigma$ are the $SU(2)$- connection $A^i_a$ and the densitized triad $\tilde{E}^b_j$, with the only nontrivial Poisson bracket $
\{A^i_a(x),\tilde{E}^b_j(y)\}=\kappa\beta\delta^3(x,y)$, where $\kappa\equiv8\pi G$ and $\beta$ is the Barbero-Immirzi parameter.  The Hamiltonian constraint reads
\begin{align}\label{class-Ham-def}
 H(N)&=\frac{1}{2\kappa}\int_{\Sigma} d^3x\;N\frac{\tilde{E}^a_i\tilde{E}^b_j}{\sqrt{\det{(q)}}}\left(\epsilon_{ijk}F^k_{ab}-2(1+\beta^2)K^i_{[a}K^j_{b]}\right)\notag\\
 &=: H^E(N)-T(N),
\end{align}
where $F^i_{ab}\equiv 2\partial_{[a}A^i_{b]}+{\epsilon^i}_{jk}A^j_aA^k_b$ is the curvature of $A^i_a$, $K^i_a$ is the extrinsic curvature of $\Sigma$, and $\det{(q)}$ is the determinate of 3-metric $q_{ab}\equiv e^i_ae^j_b\delta_{ij}$ with $e^i_a$ being the co-triad. $H^E(N)$ and $T(N)$ are called the Euclidean and Lorentzian terms of the Hamiltonian constraints respectively. Both of $H^E(N)$ and $T(N)$ depend on the canonical variables in non-polynomial ways.
Besides the indication of spin foam models, it is argued in \cite{Thiemann:2006cf} that the momentum variables in $H^E(N)$ also imply the creation of new vertices by its action. Thus we adopt the so-called semi-quantized regularization approach developed in \cite{Thiemann:1997rt} to derive a new Hamiltonian constraint operator, which creates new vertices as well. The Hamiltonian is not neglected at the co-planar vertices of spin networks by the regularization. But the result of its action on the co-planar vertices is zero. Hence it has an anomaly-free algebra on shell. 

Let us first consider $H^E(N)$. By introducing a characteristic function $\chi_\epsilon(x,y)$ such that $\lim\limits_{\epsilon\rightarrow0}\chi_\epsilon(x,y)/\epsilon^3=\delta^3(x,y)$ and using the point-splitting scheme, it can be regularized as
\begin{align}\label{Euclidean-hamiltonian-classical}
H^{E}(N)&=\frac{1}{2\kappa}\lim_{\epsilon\rightarrow0}\int_{\Sigma}{\rm d}^3x\;N(x)V_{(x,\epsilon)}^{-1/2}\epsilon_{ijk}F^i_{ab}(x)\tilde{E}^a_j(x)\notag\\
&\hspace{1.5cm}\times\int_\Sigma{\rm d}^3y\;\chi_\epsilon(x,y)\tilde{E}^b_k(y)V_{(y,\epsilon)}^{-1/2},
\end{align}
where $V_{(x,\epsilon)}:=\epsilon^3\sqrt{\det(q)}(x)$. Since the volume operator has a large kernel, the naive inverse volume operator is not well defined.  However, one can use the idea in \cite{Tikhonov:1943} to circumvent this problem by defining a permissible inverse square root of volume operator as
\begin{align}\label{inverse-volume}
\widehat{V_{(y,\epsilon)}^{-1/2}}:=\lim_{\lambda\rightarrow0}(\hat{V}_{(y,\epsilon)}+\lambda\ell^3_{\rm p})^{-1}\hat{V}^{1/2}_{(y,\epsilon)},
\end{align}
where $\hat{V}_{(y,\epsilon)}$ is the $standard$ volume operator in LQG (see \cite{Ashtekar:1997fb}) corresponding the volume of the cube with center $y$ and radial $\epsilon$. It is easy to see that qualitatively $\widehat{V^{-1/2}}$ has the same properties with $\hat{V}$. Thus we can promote the classical volume in \eqref{Euclidean-hamiltonian-classical} into its quantum version \eqref{inverse-volume} and replace both of densitized triads in \eqref{Euclidean-hamiltonian-classical} by corresponding operators $\hat{E}^b_k(y)=-i\beta\ell^2_{\rm p}\delta/\delta A^k_b(y)$ where $\ell^2_{\rm p}=\hbar\kappa$. Acting on a cylindrical function $f_\gamma$, the result formally reads
\begin{widetext}
\begin{align}\label{hamiltonian-eq:reg-integral}
&\frac{\left(-i\beta\ell^2_{\rm p}\right)^2}{2\kappa}\lim_{\epsilon\rightarrow0}\int_0^1 {\rm d}t'\int_0^1 {\rm d}t\left\{\sum_{e'\neq e}\chi_\epsilon\left(e'(t'),e(t)\right)N(e'(t'))V_{\left(e'(t'),\epsilon\right)}^{-1/2}\left[\epsilon_{ijk}F^i_{ab}(e'(t'))\dot{e}'^a(t')\dot{e}^b(t)\right]X^j_{e'}(t')X^k_e(t)\widehat{V_{\left(e(t),\epsilon\right)}^{-1/2}}\right.\notag\\
&\hspace{1cm}\left.+\sum_{e}\chi_\epsilon\left(e(t'),e(t)\right)N(e(t'))V_{\left(e(t'),\epsilon\right)}^{-1/2}\left[\epsilon_{ijk}F^i_{ab}(e(t'))\dot{e}^a(t')\dot{e}^b(t)\right]\left[\theta(t,t')X^{kj}_e(t,t')+\theta(t',t)X^{jk}_e(t',t)\right]\widehat{V_{\left(e(t),\epsilon\right)}^{-1/2}}\right\}\cdot f_\gamma,
\end{align}
where $\theta(t,t')=1$ for $t'>t$ and zero otherwise, $X^k_e(t):={\rm tr}[(h_{e(0,t)}\tau_kh_{e(t,1)})^T\partial/\partial h_{e(0,1)}]$, $X^{jk}_e(t',t):={\rm tr}[(h_{e(0,t')}\tau_jh_{e(t',t)}\tau_kh_{e(t,1)})^T\partial/\partial h_{e(0,1)}]$, here $\quad\tau_k:=-\frac{i}{2}\sigma_k$ with $\sigma_k$ being the Pauli matrices, and $T$ denotes transpose. Partitioning of the domain $[0,1]$ as $N$ segments by inputing $N-1$ points, $0=t_0, t_1, \cdots, t_{N-1}, t_N=1$, and setting $\Delta t_n\equiv t_n-t_{n-1}\equiv\delta$, the integral in \eqref{hamiltonian-eq:reg-integral} can be replaced by the Riemann's sum in a limit. Then \eqref{hamiltonian-eq:reg-integral} reduces to
\begin{align}\label{hamiltonian-eq:reg-sum}
&\frac{\left(-i\beta\ell^2_{\rm p}\right)^2}{2\kappa}\lim_{\delta\rightarrow0}\lim_{\epsilon\rightarrow0}\delta^2\left\{\sum_{e'\neq e}\sum_{n,m=1}^N\chi_\epsilon\left(e'(t'_{m-1}),e(t_{n-1})\right)N(e'(t'_{m-1}))V_{\left(e'(t'_{m-1}),\epsilon\right)}^{-1/2}\left[\epsilon_{ijk}F^i_{ab}(e'(t'_{m-1}))\dot{e}'^a(t'_{m-1})\dot{e}^b(t_{n-1})\right]\right.\notag\\
&\hspace{4cm}\times X^j_{e'}(t'_{m-1})X^k_e(t_{n-1})\widehat{V_{\left(e(t_{n-1}),\epsilon\right)}^{-1/2}}\notag\\
&\hspace{3cm}\left.+\sum_{e}\sum_{n,m=1}^N\chi_\epsilon\left(e(t_{m-1}),e(t_{n-1})\right)N(e(t_{m-1}))V_{\left(e(t_{m-1}),\epsilon\right)}^{-1/2}\left[\epsilon_{ijk}F^i_{ab}(e(t_{m-1}))\dot{e}^a(t_{m-1})\dot{e}^b(t_{n-1})\right]\right.\notag\\
&\hspace{4cm}\left.\times\left[\theta(t_{n-1},t_{m-1})X^{kj}_e(t_{n-1},t_{m-1})+\theta(t_{m-1},t_{n-1})X^{jk}_e(t_{m-1},t_{n-1})\right]\widehat{V_{\left(e(t_{n-1}),\epsilon\right)}^{-1/2}}\right\}\cdot f_\gamma.
\end{align}
\end{widetext}
Since the volume operator and hence $\widehat{V^{-1/2}_{(y,\epsilon)}}$ vanish at divalent vertices, for sufficiently small $\epsilon$, the only non-vanishing terms of the summation in \eqref{hamiltonian-eq:reg-sum} correspond to those of $m=n=1$. Moreover, the terms corresponding to $m=n=1$ in the second summation of \eqref{hamiltonian-eq:reg-sum}, which involves only summation over $e(=e')$, vanish due to $\epsilon_{ijk}F^i_{ab}(e(t'_0))\dot{e}^a(t'_0)\dot{e}^b(t_0)=0$. For $e\neq e'$, we have
\begin{align}\label{loop}
\delta^2F^i_{ab}(e'(0))\dot{e}'^a(0)\dot{e}^b(0)\approx\frac{2}{{\cal N}_{\ell}}{\rm sgn}(e',e){\rm tr}_{\ell}\left(h_{\alpha_{e'e}}\tau_i\right),
\end{align}
where ${\cal N}_{\ell}=-\frac{\ell(\ell+1)(2\ell+1)}{3}$ with $\ell$ a half-integer representing a spin representation of $SU(2)$, $\alpha_{e'e}$ is a loop formed by adding an arc between $e'(\delta)$ and $e(\delta)$, and ${\rm sgn}(e',e):={\rm sgn}[\epsilon_{ab}\dot{e}'^a(0)\dot{e}^b(0)]$ is the orientation factor which can be promoted into its quantum operator. From the property of $\hat{V}$, we know that $\widehat{V_{(v,\epsilon)}^{1/2}}\equiv\widehat{V_v^{1/2}}$ is independent of $\epsilon$. Thus we can take the trivial limit $\epsilon\rightarrow0$ and obtain
\begin{align}\label{hamiltonian-point-final}
&\hat{H}_{\delta}^{E}(N)\cdot f_\gamma:=\frac{\left(\beta\ell^2_{\rm p}\right)^2}{\kappa{\cal N}_{\ell}}\sum_{v \in V(\gamma)}N_v\widehat{V_v^{-1/2}}\notag\\
&\quad\times\left[\sum_{e\cap e'=v}{\rm sgn}(e',e)\epsilon_{ijk}{\rm tr}_{\ell}\left(h_{\alpha_{e'e}}\tau_i\right)J^j_{e'}J^k_e\right]\;\widehat{V_v^{-1/2}}\cdot f_\gamma\notag\\
&=:\sum_{v \in V(\gamma)}N_v\sum_{e\cap e'=v}\hat{H}^E_{v,e'e}
\end{align}
where $N_v\equiv N(v)$ and $J^i_e\equiv-iX^i_e$ is the self-adjoint right-invariant operator. The assignment of  $\alpha_{e'e}$ is diffeormorphism covariant in the sense that, for any $\varphi\in {\rm Diff}(\Sigma)$ there exists $\varphi'\in {\rm Diff}(\Sigma)$ such that $\varphi'(\varphi(\gamma))=\varphi(\gamma)$ and $\varphi'\left(\alpha_{\varphi(e')\varphi(e)}\right)=\varphi(\alpha_{e'e})$ \cite{Thiemann:1996aw}. Applying $\hat{H}^E_{v,e'e}$ on the $n$-valent non-planar vertex $v$ of $T^v_{\gamma,\vec{j},\vec{i}}$\,, the intertwiner $i_v$ associated to $v$ will be changed to $i'_v$, while, as a multiplication operator,  $h_{\alpha_{e'e}}$ will change the spins associated to the segments of $e'$ and $e$. Hence the action of $\hat{H}^E_{v,e'e}$ is given by
\begin{align}\label{action-of-HE}
\makeSymbol{
\includegraphics[width=2.9in]{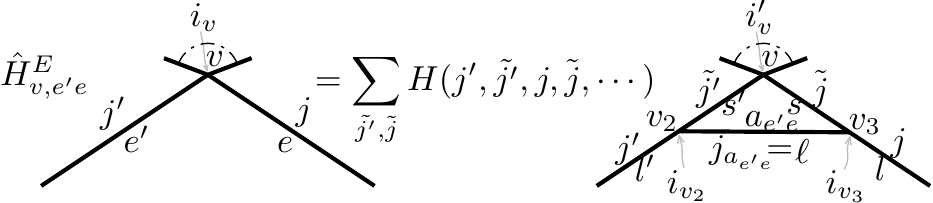}}\,,
\end{align}
where $\tilde{j'}\in \{|j'-\ell|,\cdots,j'+\ell\},\tilde{j}\in\{|j-\ell|,\cdots,j+\ell\}$, and $H(j',\tilde{j'},j,\tilde{j},\cdots)$ are coefficients with $``\cdots"$ denoting spins associated to other edges incident at $v$.

The above regularization approach can be similarly used to quantize the Lorentzian term in the Hamiltonian \eqref{class-Ham-def}. The Lorentzian term can be regularized as 
\begin{align}\label{reg-Loren}
T(N)&=\frac{2(1+\beta^2)}{2\kappa}\lim_{\epsilon\rightarrow0}\int_{\Sigma}{\rm d}^3x\;N(x)V_{(x,\epsilon)}^{-1/2}(K^{[j}_aK^{k]}_b)(x)\tilde{E}^a_j(x)\notag\\
&\hspace{2cm}\times\int_\Sigma{\rm d}^3y\;\chi_\epsilon(x,y)\tilde{E}^b_k(y)V_{(y,\epsilon)}^{-1/2}.
\end{align}
Following the regularization procedure of the Euclidean term, Eq. \eqref{loop} is now replaced by
\begin{align}
\delta^2K^{[j}_a(e'(0))K^{k]}_b(e'(0))\dot{e}'^a(0)\dot{e}^b(0)\approx{\rm sgn}(e',e)K^j_{e'}K^k_e,
\end{align}
where $K^j_{e'}:=-\frac{1}{\kappa\beta{{\cal N}_{\ell}}}\left(\tau_jh_{s_{e'}}\left\{h_{s_{e'}}^{-1},\bar{K}\right\}\right)$,
with $s_{e'}$ as the starting segment of $e'$ with parameter length $\delta$, and we have used the identity
$K^j_a=\frac{1}{\kappa\beta}\{A^j_a,\bar{K}\}=\frac{1}{\kappa\beta{\cal N}_\ell}{\rm tr}_\ell(\tau_j\{A_a,\bar{K}\})$,
with $\bar{K}:=\int_\Sigma {\rm d}^3x \,K^i_a\tilde{E}^a_i=\frac{1}{\beta^2}\{H^E(1),V\}$ \cite{Thiemann:1996aw}. Replacing the Poisson brackets by commutators times $1/(i\hbar)$, we obtain the quantized Lorentzian term
\begin{align}\label{hamiltonian-point-final-Loren}
\hat{T}_\delta(N)\cdot f_\gamma&=\frac{2(1+\beta^2)\left(\beta\ell_{\rm p}^2\right)^2}{2\kappa}\sum_{v\in V(\gamma)}N_v\widehat{V_v^{-1/2}}\notag\\
&\quad\times\left[\sum_{e\cap e'=v}{\rm sgn}(e',e)\hat{K}^j_{e'}\hat{K}^k_{e}J^j_{e'}J^k_e\right]\widehat{V_v^{-1/2}}\cdot f_\gamma,
\end{align}
where
\begin{align}
\hat{K}^i_e&:=-\frac{1}{i\ell^2_{\rm p}\beta{{\cal N}_{\ell}}}\left(\tau_ih_{s_e}\left[h_{s_e}^{-1},\hat{\bar{K}}\right]\right),\text{with}\;
\hat{\bar{K}}:=\frac{1}{i\hbar\beta^2}\left[\hat{H}^E_\delta(1),\hat{V}\right].
\end{align}
Hence the total regulated Hamiltonian constraint operator is given by
\begin{align}\label{total-Ham}
\hat{H}_\delta(N)\cdot f_\gamma&:=\sum_{v \in V(\gamma)}N_v\left(\hat{H}^E_v-\hat{T}_v\right)\cdot f_\gamma=:\sum_{v \in V(\gamma)}N_v\hat{H}_v\cdot f_\gamma.
\end{align}
Notice that for a given $\gamma$, there are indeed finite terms contribution to the summation no matter how fine the partition is. Hence the operator \eqref{total-Ham} is well defined in ${\cal H}_{\rm kin}$. Since the volume operator and hence $\widehat{V^{-1/2}_{(y,\epsilon)}}$ vanish at internal gauge invariant trivalent vertices as well, $\hat{H}_\delta(N)$ acts nontrivially only on non-planar vertices with valence higher than three. Note that we can also employ a fourfold point-splitting scheme as in \cite{Thiemann:1997rt} to regulate the Hamiltonian constraint. In this case $1/\sqrt{\det(q)}$ in \eqref{class-Ham-def}  can be absorbed into Poisson brackets, and the final operator takes the same form as \eqref{hamiltonian-point-final} and \eqref{hamiltonian-point-final-Loren}, except that $\widehat{V_v^{-1/2}}$ is replaced by
\begin{align}
&\widehat{V_{{\rm alt},v}^{-1/2}}:=\left|-\frac{4\times8}{3!E(v)}\sum_{s_I\cap s_J\cap s_K=v}\epsilon^{IJK}{\rm tr}\left({}^{(\frac12)}\!\hat{e}_I{}^{(\frac12)}\!\hat{e}_J{}^{(\frac12)}\!\hat{e}_K\right)\right|,
\end{align}
where ${}^{(\frac12)}\!\hat{e}_I:= -\frac{2}{i\beta\ell_{\rm p}^2}h_{s_I}[h_{s_I}^{-1},\hat{V}^{1/2}_v]$. It can be shown that qualitatively $\widehat{V_{{\rm alt},v}^{-1/2}}$ has the same key properties with $\widehat{V_v^{-1/2}}$ \cite{graph-III}.

Let $\Phi:={\rm Cyl}^{\infty}(\overline{{\cal A}/{\cal G}})$ be the set of smooth cylindrical function and $\Phi'_{\rm Diff}$ be the space of diffeomorphism-invariant distributions on $\Phi$. Then the number $\psi(\hat{H}_\delta(N)\cdot f)$ for any $f\in\Phi$ and $\psi\in\Phi'_{\rm Diff}$ depends only on the diffeomorphism class of the loop assignments. Hence the limit $\delta\rightarrow0$ can be taken in the Rovelli-Smolin topology as \cite{Rovelli:1993bm,Thiemann:1996av}
\begin{align}
\left((\hat{H}(N))'\cdot \psi\right)(f_\gamma)&:=\lim_{\delta\rightarrow0}\left((\hat{H}_\delta(N))'\cdot\psi\right)(f_\gamma)\notag\\
&:=\lim_{\delta\rightarrow0}\psi(\hat{H}_\delta(N)\cdot f_\gamma).
\end{align}
However, $(\hat{H}(N))'$ is not well defined in  $\Phi'_{\rm Diff}$, since it does not keep  $\Phi'_{\rm Diff}$ invariant. Taking account of the fact that the assignment of loops by $\hat{H}_\delta(N)$ is diffeomorphism covariant and it acts nontrivially only on non-planar vertices with valence higher than three, we consider almost diffeomorphism invariant states which are obtained from the spin network states by averaging over their images under diffeomorphisms but leaving fixed sets of non-planar vertices with valence higher than $3$ in the spatial manifold invariant, parallel to the proposal in \cite{Lewandowski:2014hza}. Given a graph $\gamma$, we denote its non-planar vertices with valence higher than $3$ by $V_{\rm np4}(\gamma)$, the group of all diffeomorphisms preserving $V_{\rm np4}(\gamma)$ by ${\rm Diff}(\Sigma)_{V_{\rm np4}(\gamma)}$, and the diffeomorphism acting trivially on $\gamma$ by ${\rm TDiff}(\Sigma)_\gamma$.  For any $f_\gamma\in \Phi$, we define a map $\eta:\Phi\rightarrow \Phi'$ by
\begin{align}
\eta(f_\gamma):=\frac{1}{N_\gamma}\sum_{\varphi\in {\rm Diff}(\Sigma)_{V_{\rm np4}(\gamma)}/{\rm TDiff}(\Sigma)_\gamma}\hat{U}_\varphi\cdot f_\gamma,
\end{align}
where $\hat{U}_\varphi$ is the unitary representation of $\varphi$, and $N_\gamma$ is a normalization factor. We can equip  the space $\Phi_{\rm np4}:=\eta(\Phi)$ with a natural inner product as $\langle\eta(f)|\eta(g)\rangle:=\eta(f)(g), \forall f,g\in \Phi$. The new Hilbert space ${\cal H}_{\rm np4}$ is defined as the completion of $\Phi_{\rm np4}$. Note that $\Phi_{\rm np4}$ is also in the dual space of the space of diffeomorphism invariant states up to all vertices of spin networks which was fully discussed in \cite{Lewandowski:2014hza}. Whether $(\hat{H}(N))'$ can be well defined in ${\cal H}_{\rm np4}$ is a delicate issue. In particular, if the action of $\hat{H}_\delta(N)$ on a non-planar vertex with higher valence could deduce the valence so that it becomes less than $4$ or the vertex becomes co-planar, $(\hat{H}(N))'$ would be ill-defined on the resulted almost diffeomorphism invariant states.\footnote{Thanks to the comment from Jerzy Lewandowski.} Fortunately, we can use the freedom of choosing the spin representations $\ell$ attached to each new added loop in \eqref{loop} to ensure that the valence of any vertex would not be changed by the action of $\hat{H}_\delta(N)$. Then it is straightforward to see that $(\hat{H}(N))'$ is well defined in ${\cal H}_{\rm np4}$. By the dual action, $(\hat{H}(N))'$ will annihilate the arcs like the one connecting $v_2$ and $v_3$ in \eqref{action-of-HE}. Since $\hat{H}_\delta(N)$ vanishes at co-planar vertices, it is obvious that
\begin{align}
&\left([(\hat{H}(M))',(\hat{H}(N))']\cdot \phi\right)(f_\gamma)\notag\\
=&\phi\left(\left(\hat{H}_{\delta'}(N)\hat{H}_\delta(M)-\hat{H}_{\delta'}(M)\hat{H}_\delta(N)\right)\cdot f_\gamma\right)\notag\\
=&0,\quad\quad \forall f_\gamma\in\Phi,\phi\in\Phi_{\rm np4}.
\end{align}
Hence for any $\psi\in\Phi'_{\rm Diff}$, which is also a distribution on $\Phi_{\rm np4}$, we have
\begin{align}
\left(\left([(\hat{H}(M))',(\hat{H}(N))']\right)'\cdot\psi\right)(\phi)&=\psi\left([(\hat{H}(M))',(\hat{H}(M))')]\cdot\phi\right)\notag\\
&=0,\quad\quad\forall\phi\in\Phi_{\rm np4}.
\end{align}
Therefore the quantum algebra of the Hamiltonian constraint operators $(\hat{H}(N))'$ is anomaly-free on shell. A symmetric Hamiltonian constraint operator corresponding to $(\hat{H}(N))'$ can be defined in ${\cal H}_{\rm np4}$ as
\begin{align}\label{Ham-sym}
(\hat{H}^{\rm sym}(N))':=\frac{1}{2}\left((\hat{H}(N))'+((\hat{H}(N))')^\dag\right), 
\end{align}
where $((\hat{H}(N))')^\dag$ denotes the adjoint of $(\hat{H}(N))'$. The action of $((\hat{H}(N))')^\dag$ on an almost diffeomorphism invariant state will create co-planar vertices and arcs like in \eqref{action-of-HE}. Thus it is easy to see that the quantum algebra of $(\hat{H}^{\rm sym}(N))'$ is also anomaly-free.

Now we test the above regularization technique of the anomaly-free Hamiltonian constraint operators in the symmetry-reduced model of LQC. For simplification, we consider only the spatially flat and isotropic model. The construction can be generalized to the other cosmological models directly. Introducing an elementary cell ${\cal V}$ adapted to the fiducial triad, the connections $A_a^i$ and the densitized triads $E^a_i$ can be expressed as $A_a^i = cV_o^{-1/3}\,{}^o\!e_a^i$ and
$E^a_i = pV_o^{-2/3}\sqrt{{}^o\!q}\, {}^o\!e^a_i$, where $({}^o\!e_a^i, {}^o\!e^a_i)$ are a set of orthonormal co-triads and triads compatible with the fiducial metric ${{}^o\!q}_{ab}$ \cite{Ashtekar:2003hd}. The basic (nonvanishing) Poisson bracket is given by $\{c,\, p\} = \frac{\kappa\beta}{3}$. To pass to the quantum theory, one constructs a kinematical Hilbert space ${\cal H}^{\mathrm{grav}}_{\mathrm{kin}}=L^2(\mathbb{R}_{\mathrm{Bohr}},{\mathrm{d}}\mu_{\mathrm{Bohr}})$, where $\mathbb{R}_{\mathrm{Bohr}}$ is the Bohr compactification of the real line and ${\mathrm{d}}\mu_{\mathrm{Bohr}}$ is the Haar measure on it \cite{Ashtekar:2003hd}. The holonomy of $A^i_a$ along an edge, parallel to the triad vector ${}^o\!e^a_i$, of length $\bar{\mu}V_o^{1/3}$ with respect to ${{}^o\!q}_{ab}$ is given by $h^{\bar{\mu}}_i=\cos\frac{\bar{\mu}c}{2}\mathbb{I}+2\sin\frac{\bar{\mu}c}{2}\tau_i$, where $\mathbb{I}$ is the identity $2\times2$ matrix. Because of homogeneity, the Hamiltonian constraint can be written as \cite{Ashtekar:2003hd,Ashtekar:2006wn}
\begin{align}\label{full-H-cosmology}
H&=-\frac{1}{\beta^2}H^E(1)=-\frac{1}{2\kappa\beta^2}\int_{\cal V}d^3x\;\frac{\epsilon_{ijk}F^i_{ab}\tilde{E}^a_j\tilde{E}^b_k}{\sqrt{\det(q)}}.
\end{align}
Since $\sqrt{\det(q)}=V\left(V_o^{-1}\sqrt{\det({}^o\!q)}\right)$ where $V:=|p|^{3/2}$ is the physical volume of ${\cal V}$, we can write $H$ as
\begin{align}
H&=-\frac{1}{2\kappa\beta^2}\int_{\cal V}d^3x\;\frac{V_o}{\sqrt{\det({}^o\!q)}}\;{V^{-1/2}}\epsilon_{ijk}F^i_{ab}\tilde{E}^a_j\tilde{E}^b_kV^{-1/2}.
\end{align}
Note that $\tilde{E}^a_j$ can be directly quantized as
\begin{align}
\hat{\tilde{E}}^a_j=V_o^{-2/3}\sqrt{\det({}^o\!q)}\;{}^o\!e^a_j\hat{p}:=V_o^{-2/3}\sqrt{\det({}^o\!q)}\;{}^o\!e^a_j\times\left(-i\hbar\frac{\kappa\beta}{3}\right)\frac{{\rm d}}{{\rm d}c}\,.
\end{align}
After the action of $\hat{\tilde{E}}^a_j$ and $\hat{\tilde{E}}^b_k$, we can integrate $F^i_{ab}$ over the plane spanned by ${}^o\!e^a_j$ and ${}^o\!e^b_k$ to yield a holonomy along a loop. Promoting functions into corresponding operators, the  regulated operator corresponding to \eqref{full-H-cosmology} reads
\begin{align}
\hat{H}^{\bar{\mu}}=-\frac{1}{2\kappa\beta^2{\cal N}_{\ell}}\widehat{V^{-1/2}}\frac{\epsilon_{ijk}{\rm tr}_{\ell}\left(\hat{h}_{\alpha^{\bar{\mu}}_{jk}}\tau_i\right)}{\bar{\mu}^2}\hat{p}\hat{p}\widehat{V^{-1/2}}\ ,
\end{align}
where $\alpha^{\bar{\mu}}_{jk}$ is a loop along a square in the $j$-$k$ plane spanned by a face of ${\cal V}$, each of whose sides has length $\bar{\mu}V_o^{1/3}$ with respect to ${}^o\!q_{ab}$, and $h_{\alpha^{\bar{\mu}}_{jk}}=h^{\bar{\mu}}_jh^{\bar{\mu}}_k\left(h^{\bar{\mu}}_j\right)^{-1}\left(h^{\bar{\mu}}_k\right)^{-1}$.
Since the area operator in LQG has a minimum nonzero eigenvalue $2\sqrt{3}\pi\beta\ell^2_{\rm p}$ which introduces a nature cut off to the size of the loop $\alpha^{\bar{\mu}}_{jk}$, we obtain
\begin{align}\label{LQC-ham}
\hat{H}=-\frac{3}{\kappa\beta^2}\widehat{V^{-1/2}}\frac{\sin^2({\bar{\mu}c})}{\bar{\mu}^2}\hat{p}\hat{p}\widehat{V^{-1/2}}, \quad\bar{\mu}^2|p|=2\sqrt{3}\pi\beta\ell^2_{\rm p}.
\end{align}
The symmetric Hamiltonian constraint operator can be defined by
\begin{align}\label{H-cosm-sym}
\hat{H}^{\rm sym}:=\frac12(\hat{H}+\hat{H}^\dag).
\end{align}
This Hamiltonian constraint operator is obviously simpler than those appeared in LQC models (e.g. the APS Hamiltonian operator \cite{Ashtekar:2006wn}). It is straightforward to check that the Hamiltonian operator \eqref{H-cosm-sym} has correct classical limit and the classical big bang singularity can also be avoided by a quantum bounce of this dynamics.

To summarize, the Hamiltonian constraint of GR is successfully quantized in LQG by adopting the semi-quantized regularization approach. The resulted operator is symmetric and well defined in the Hilbert space ${\cal H}_{\rm np4}$ of diffeomorphism invariant states up to non-planar vertices with valences higher than $3$. The action of this Hamiltonian constraint operator creates new trivalent co-planar vertices to the spin networks but does not change the valence of the acted vertices, and hence it can be symmetric and match the quantum dynamics of spin foam models. Meanwhile, our regularization procedure ensures that there is less ambiguity in the construction of the operator. The Hamiltonian is not neglected at the co-planar vertices of spin networks by the regularization. But the result of its action on the co-planar vertices is zero. Hence the quantum algebra of the new Hamiltonian constraint operator is anomaly-free on shell. It is thus possible to find solutions of the quantum Hamiltonian constraint in ${\cal H}_{\rm np4}$. The regularization procedure for the Hamiltonian operator in full theory is also applied to the isotropic model of LQC. The resulted Hamiltonian constraint operator is simpler than those appeared in LQC models. Meanwhile, it inherits the qualitative properties of the previous Hamiltonian operators in LQC, so that it has correct classical limit and the classical big bang singularity can also be avoided by a quantum bounce.

The authors would like to thank Jerzy Lewandowski, Chopin Soo, Thomas Thiemann and Hoi-Lai Yu for useful discussions. J. Y. is supported in part by NSFC No. 11347006, and by the Institute of Physics, Academia Sinica, Taiwan. Y. M. is supported in part by the NSFC (Grant Nos. 11235003 and 11475023) and the Research Fund for the Doctoral Program of Higher Education of China.

\providecommand{\href}[2]{#2}\begingroup\raggedright\endgroup


\end{document}